\documentclass[epj]{svjour}
\usepackage{amsfonts}
\usepackage{amsmath}
\usepackage{amssymb}
\usepackage{graphics}
\begin{document}
\title{Effects of the third-order dispersion on continuous waves in complex potentials}
\author{Bin Liu\inst{1}, Lu Li\inst{1}
\thanks{e-mail: llz@sxu.edu.cn} \and  Boris A. Malomed\inst{2,3}
}
%
\date{Received: date / Revised version: date}
\institute{Institute of Theoretical Physics, Shanxi University, Taiyuan 030006, China \and Department of Physical Electronics, School of Electrical Engineering,
Faculty of Engineering, Tel Aviv University, Tel Aviv 69978, Israel \and Laboratory of Nonlinear-Optical Informatics, ITMO University, St.
Petersburg 197101, Russia}

\abstract{
A class of constant-amplitude (CA) solutions of the nonlinear Schr\"{o}dinger
equation with the third-order spatial dispersion (TOD) and complex potentials
are considered. The system can be implemented in specially designed planar
nonlinear optical waveguides carrying a distribution of local gain and loss
elements, in a combination with a photonic-crystal structure. The complex
potential is built as a solution of the inverse problem, which predicts the
potential supporting a required phase-gradient structure of the CA state. It
is shown that the diffraction of truncated CA states with a correct phase structure can
be strongly suppressed. The main subject of the analysis is the modulational
instability (MI) of the CA states. The results show that the TOD term tends to
attenuate the MI. In particular, simulations demonstrate a phenomenon of weak
stability, which occurs when the linear-stability analysis predicts small
values of the MI growth rate. The stability of the zero state, which is a nontrivial
issue in the framework of the present model, is studied too.
\PACS{ {05.45.Yv}{Solitons}
      \and {42.81.Dp}{solitons in optics}
      \and {42.81.Qb}{Fiber waveguides, couplers, and
arrays}
      \and {42.65.Tg}{Optical solitons; nonlinear guided
waves}
     } 
} 
\maketitle
\section{Introduction}

In many systems governed by wave equations, complex (non-Hermitian)
potentials give rise to effects that cannot be realized with real
(Hermitian) potentials \cite{Rep.Prog.Phys70_947}, a well-known example
being the parity-time ($\mathcal{PT}$) symmetry maintained by potentials
with spatially even real and odd imaginary parts \cite%
{PRL80_5243,JMP43_2814,PRL89_270401}. In optics, $\mathcal{PT}$ symmetric
potentials have been created in the experiments \cite%
{Guo,NP6_192,Nat488_167,NatMater12_108,NP10_394}. They exhibit remarkable
properties and potential applications, such as power oscillations \cite%
{NP6_192}, non-reciprocal light propagation \cite{PRL100_103904}, optical
transparency \cite{PRL103_093902}, unidirectional invisibility induced by
the balanced gain-loss profile \cite%
{PRA82_043803,PRL106_213901,PRL110_234101}, and $\mathcal{PT}$-symmetric
devices \cite{PRL112_143903,Science346_972,Science346_975}.\emph{\ }Optics
provides an especially fertile ground for the investigation of $PT$%
-symmetric beam dynamics in nonlinear regimes, including the formation of
bright and dark solitons, gap solitons, and vortices \cite%
{18,19,20,21,22,23,24,25,26,27,28,29,30,31,32,RMP,EPJD66_266(2012),EPJD68_322(2014),EPJD69_31(2015),EPJD70_14(2016),EPJD69_171(2015),rrp67_802,rjp61_577,rjp61_1028}%
. The concept of the $PT$-symmetry has also been applied to Bose-Einstein
condensates \cite{PRA86_013612,PRA90_042123,PRA93_033617}, atomic cells \cite%
{OL38_4033,OL39_5387} and atomic Bose-Josephson junction \cite%
{EPJD70_157(2016)}.

Recently, stability of optical solitons and nonlinear beam dynamics near
phase-transition points in non-$\mathcal{PT}$ symmetric complex potentials
(i.e., more general ones) was addressed too, cf. an earlier work on gap
solitons in the complex Ginzburg-Landau equation \cite{SH}. The results show
that the solitons may be stable in a wide range of parameter values both
below and above the phase transition \cite%
{RMP,OL41_2747,PhysicaD331_48,arXiv:1605.02259}. Some applications, such as
coherent perfect absorbers and time-reversal lasers \cite%
{PRL105_053901,Science331_889,PRL108_173901,NC5_4034,science346_328} have
been elaborated in such settings.

The simplest solutions of wave equations are continuous waves (CWs), alias
plane waves, which keep a constant amplitude propagating in the free space.
In the framework of wave equations corresponding to Hermitian Hamiltonians
with self-focusing nonlinearity, CWs are subject to the modulation
instability (MI), which was first identified in fluid mechanics \cite%
{J.FluidMech27_417} and plasma physics \cite{PRL21_209}, and subsequently
reported in many other fields \cite%
{EPJD19_223(2002),EPJD59_223(2010),EPJB74_151(2010),EPJD11_301(2000)},
in particular, in nonlinear optics \cite%
{PRL56_135,Agrawal,PRL92_163902,PRL96_014503,PhysicaD313_26,EPJB50_321(2006),EPJST203_217(2012)}%
, including the MI of CW states in two-component systems \cite{Agrawal,Rich}%
. The latter topic was recently extended by the analysis of the MI in the
spin-orbit-coupled system \cite{Ponz}.

The MI of waves in the $\mathcal{PT}$-symmetric nonlinear Schr\"{o}-dinger
(NLS) equation has been widely studied too \cite%
{arXiv:1605.02259,PRE83_036608,OL36_4323,J.Opt15_064010,oe22_19774,PRE91_023203,PRE92_022913}%
. As an extension of the studies in this direction, constant-amplitude (CA)
waves have recently been addressed in models with more general complex
potentials \cite{NatureCommunication6_7257}. In the present work, we aim to
study the MI of CA solutions of the NLS equation with complex potentials and
spatial third-order dispersion (TOD). Higher-order spatial dispersion (i.e.,
diffraction) may be engineered in photonic-crystal media, odd-order terms
appearing in the case of an oblique propagation of optical beams \cite%
{phot-cryst}. The latter term was recently added to nonlinear $\mathcal{PT}$%
-symmetric systems in Ref. \cite{ScientificReports6_23478}.

The paper is organized as follows. In the next section, the model and its
reduction are introduced, and the corresponding CA solutions are presented.
We consider three kinds of potentials, namely, localized $\mathcal{PT}$%
-symmetric, non-$\mathcal{PT}$-symmetric, and periodic $\mathcal{PT}$%
-symmetric ones. In fact, the complex potentials are built as solutions of
the inverse problem, corresponding to the phase-gradient field of the
required solution. In Sec. III, we focus on the MI of CA waves by making use
of the plane-wave-expansion method, and present dependence of the MI growth
rate on the wavenumber. Results for the stability of the zero state, which
is a necessary part of the analysis too, are also reported in Sec. III. The paper is concluded by Sec. IV.

\section{The model and constant-amplitude (CA) solutions}

We begin the analysis by considering the NLS equation with the TOD and a
complex potential, written in a scaled form, cf. Ref.
\cite{ScientificReports6_23478}:
\begin{equation}
i\frac{\partial\Psi}{\partial z}+\frac{1}{2}\frac{\partial^{2}\Psi}{\partial
x^{2}}+i\frac{\beta}{6}\frac{\partial^{3}\Psi}{\partial x^{3}}+V(x)\Psi
+g\left\vert \Psi\right\vert ^{2}\Psi=0. \label{Model}%
\end{equation}
In the application to light propagation in planar waveguides, $\Psi(x,z)$ is
the slowly varying envelope of the electric field, $z$ and $x$ are the
propagation distance and the transverse coordinate, $\beta$ is the TOD
strength, and $V(x)\equiv V_{R}(x)+iV_{I}(x)$ represents the complex
potential, which can be implemented in optics by combining the spatially
modulated refractive index and spatially distributed loss and gain elements
\cite{NP6_192}. The nonlinearity may be either self-focusing (with $g>0$) or
defocusing, with $g<0$, the latter being possible in semiconductor materials
\cite{semicond}.

As mentioned above, the TOD term in the spatial domain may appear in the
waveguide carrying a photonic-crystal structure, which can modify the simple
paraxial form of the diffraction. Then, if the carrier beam is tilted with
respect to the underlying structure, the effective diffraction operator in the
propagation equation will acquire the TOD term, similar to its counterpart
appearing in the temporal domain for narrow wave packets \cite{Agrawal}.

We are looking for stationary solutions of Eq. (\ref{Model}) as%
\begin{equation}
\Psi(x,z)=\Phi(x)\exp(i\mu z), \label{Solution1}%
\end{equation}
where $\mu$ is a real propagation constant, and complex field profile
$\Phi(x)$ obeys the following nonlinear equation:
\begin{equation}
-\mu\Phi+\frac{1}{2}\Phi^{\prime\prime}+\frac{i\beta}{6}\Phi^{\prime
\prime\prime}+V(x)\Phi+g\left\vert \Phi\right\vert ^{2}\Phi=0, \label{Model2}%
\end{equation}
with the prime standing for $d/dx$. Further, we define real amplitude and
phase
\begin{equation}
\Phi(x)=H(x)\exp\left[  i\Theta(x)\right]  , \label{Solution2}%
\end{equation}
for which complex equation (\ref{Model2}) splits into real ones:%
\begin{align}
0  &  =-\mu H+\frac{1}{2}H^{\prime\prime}-\frac{1}{2}H\left(  \Theta^{\prime
}\right)  ^{2}+\frac{\beta}{6}H\left(  \Theta^{\prime}\right)  ^{3}%
-\frac{\beta}{6}H\Theta^{\prime\prime\prime}\nonumber\\
&  +V_{R}H+g\left\vert H\right\vert ^{2}H-\frac{\beta}{2}H^{\prime\prime
}\Theta^{\prime}-\frac{\beta}{2}H^{\prime}\Theta^{\prime\prime}%
,\label{Real part}\\
0  &  =H^{\prime}\Theta^{\prime}+\frac{1}{2}H\Theta^{\prime\prime}+\frac
{\beta}{6}H^{\prime\prime\prime}-\frac{\beta}{2}H^{\prime}\left(
\Theta^{\prime}\right)  ^{2}\nonumber\\
&  -\frac{\beta}{2}H\Theta^{\prime}\Theta^{\prime\prime}+V_{I}H.
\label{Imag part}%
\end{align}

Assuming a CA solution, namely $H(x)=A=\mathrm{const}$, and choosing the
propagation constant as $\mu=gA^{2}$, Eqs. (\ref{Real part}) and
(\ref{Imag part}) amount to the following expression for the complex
potential, which actually produces a solution of the \textit{inverse problem}
in the present context (selection of potentials which support a required form
of the solution, see, e.g., Ref. \cite{Stepa}):
\begin{align}
V_{R}\left(  x\right)   &  =\frac{\beta}{6}\Theta^{\prime\prime\prime}%
+\frac{1}{2}\left(  \Theta^{\prime}\right)  ^{2}-\frac{\beta}{6}\left(
\Theta^{\prime}\right)  ^{3},\label{R}\\
V_{I}\left(  x\right)   &  =\frac{\beta}{2}\Theta^{\prime}\Theta^{\prime
\prime}-\frac{1}{2}\Theta^{\prime\prime}. \label{I}%
\end{align}
Accordingly, the CA solution with any relevant real-valued phase function
$\Theta(x)$ is constructed as%
\begin{equation}
\Psi(x,z)=A\exp\left[  igA^{2}z+i\Theta(x)\right]  , \label{solution3}%
\end{equation}
if the complex potential is chosen as per Eqs. (\ref{R}) and (\ref{I}).
Further, setting $\Theta^{\prime}(x)\equiv W(x)$, the CA solution is written
as%
\begin{equation}
\Psi(x,z)=A\exp\left[  igA^{2}z+i%
{\displaystyle\int}
W\left(  x\right)  dx\right]  , \label{solution}%
\end{equation}
with the real and imaginary parts of the complex potential being%
\begin{align}
V_{R}\left(  x\right)   &  =\frac{\beta}{6}W^{\prime\prime}+\frac{1}{2}%
W^{2}-\frac{\beta}{6}W^{3},\label{VR}\\
V_{I}\left(  x\right)   &  =\frac{\beta}{2}WW^{\prime}-\frac{1}{2}W^{\prime},
\label{VI}%
\end{align}
where $W(x)$ is any real-valued function, and $A$ a real constant, which we
define to be positive, without the loss of generality.

If $\beta=0$, CA solution (\ref{solution}) reduces to one for the usual NLS
equation with the complex potential, $V(x)=W^{2}/2-iW_{x}/2$, which was considered in Ref.
\cite{NatureCommunication6_7257}, and is often called the Wadati potential \cite{VVK,VVK1}. Here we address a more general situation,
including the TOD effect, which affects the complex potential given by Eqs.
(\ref{VR}) and (\ref{VI}). Furthermore, it follows from Eqs. (\ref{VR}) and
(\ref{VI}) that, if $W(x)$ is an even function of $x$, complex potential
$V(x)$ is $\mathcal{PT}$-symmetric. However, the CA solution (\ref{solution})
is valid for any real-valued function $W(x)$, which implies that complex
potential $V(x)$ does not need to be a $\mathcal{PT}$-symmetric one. It should
be noted that the CA solution exists in the linear limit ($g=0$), as well as
for an arbitrary strength of the nonlinearity ($g\neq0$). It can also be shown
that the real-valued function $W(x)$ determines the power flow from the gain
to the loss regions, the respective Poynting vector, $S=(i/2)(\Psi\partial
\Psi^{\ast}/\partial x-\Psi^{\ast}\partial\Psi/\partial x)$, taking a very
simple form, $S=A^{2}W$

\begin{figure}[tbp]
\resizebox{0.5\textwidth}{!}{  \includegraphics{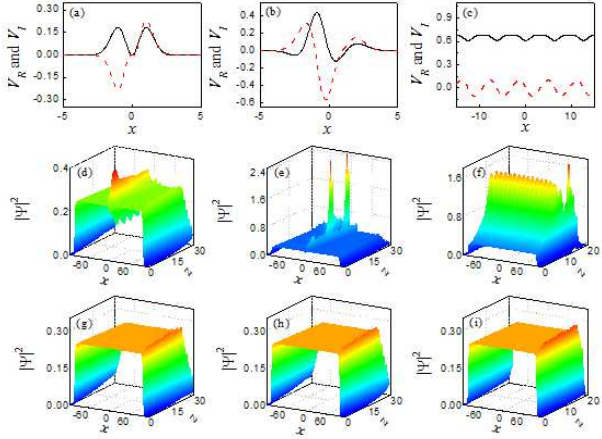}
} 
\caption{ Real and imaginary parts of complex potential (\protect\ref{VR})
and (\protect\ref{VI}) are shown by the black solid and red dashed curves,
respectively, for (a) $W(x)=\exp(-B_{1}x^{2})$, (b) $W(x)=x\exp(-B_{2}x^{2})$%
, and (c) $W(x)=V_{0}/2+V_{1}\cos(qx)$. (d-f): Evolution of truncated CA
solutions which are not supported by the correct phase structure, in the
presence of the potentials corresponding to (a-c). (g-i): The evolution of
the truncated solutions with the correct phase structure in the same cases (a-c). Here
we set $A=0.5$ and $\protect\beta=1$, the other parameters being $B_{1}=1$
in (a); $B_{2}=0.5$ in (b); $V_{0}=4$, $V_{1}=0.2$ and $q=1$ in (c); $%
B_{1}=1 $ and $g=-1$ in (d) and (g); $B_{2}=0.5$ and $g=-1$ in (e) and (h);
and $V_{0}=4$, $V_{1}=0.2$, $q=1$ and $g=1$ in (f) and (i).}
\label{fig:1}
\end{figure}

To exhibit properties of such CA solutions, we consider three complex
potentials which are relevant for various physical realizations,
\textit{viz}., ones generated by $W(x)=\exp(-B_{1}x^{2})$, $W(x)=x\exp
(-B_{2}x^{2})$, and
\begin{equation}
W(x)=\frac{1}{2}V_{0}+V_{1}\cos(qx). \label{W}%
\end{equation}
The first and last potentials are $\mathcal{PT}$-symmetric, while the second
one is not, as shown in Figs. 1(a)-1(c).

To illustrate the dynamics of CA states under the action of these potentials,
we have simulated the evolution of such states in a spatially truncated form.
When the initial state is not given the correct phase structure, determined by solution
(\ref{solution}), but is simply taken in a real form, $\Psi(x,0)\equiv A$, the
light diffracts fast to the gain region, as seen in Figs. 1(d)-1(f). On the
other hand, for the truncated solution carrying the correct phase distribution, the
diffraction is strongly suppressed, as shown in Figs. 1(g)-1(i). We also find
that, naturally, the larger the truncation width is at $z=0$, the longer the
diffraction-free propagation distance is (not shown here in detail). In the
rest of the paper, we focus on CA states supported by the periodic complex
potential (\ref{W}), which is most relevant for the realization by means of
photonic lattices in optical media.

\section{Modulational instability of the constant-amplitude (CA) solutions}

In this section, we address the stability of the CA solutions by employing the
linear-stability analysis, based on the plane-wave-expansion method, and
direct numerical simulations. We note that, although the familiar
linear-stability analysis, previously elaborated for models with localized potentials
$V(x)$ \cite{BOOK_YJK}, can be readily applied to Eq. (\ref{Model}), it cannot
be used to obtain the stability band structure in the presence of the periodic
potential. Here, we apply the plane-wave-expansion method
\cite{NatureCommunication6_7257} to study the stability of the CA solution
(\ref{solution}) in the latter case. As an example of the periodic potential,
we choose the $\mathcal{PT}$-symmetric one, which is displayed in Fig. 1(c)
and generated by Eqs. (\ref{VR}) and (\ref{VI}) with $W(x)=V_{0}/2+V_{1}%
\cos(qx)$:%
\begin{align}
V_{R}\left(  x\right)   &  =-\frac{\beta V_{1}q^{2}}{6}\cos\left(  qx\right)
+\frac{1}{2}\left[  \frac{V_{0}}{2}+V_{1}\cos\left(  qx\right)  \right]
^{2}\nonumber\\
&  -\frac{\beta}{6}\left[  \frac{V_{0}}{2}+V_{1}\cos\left(  qx\right)
\right]  ^{3},\label{Vr}\\
V_{I}\left(  x\right)   &  =\frac{qV_{1}}{2}\left[  1-\beta\left(  \frac
{V_{0}}{2}+V_{1}\cos\left(  qx\right)  \right)  \right]  \sin\left(
qx\right)  , \label{Vi}%
\end{align}
the corresponding CA solution (\ref{solution}) being
\begin{equation}
\Psi\left(  x,z\right)  =A\exp\left[  i\Xi\left(  x,z\right)  \right]  ,
\label{Psi}%
\end{equation}
where $\Xi\left(  x,z\right)  =gA^{2}z+V_{0}x/2+(V_{1}/q)\sin(qx)$.

The linear-stability analysis can be performed by adding small perturbations
to the CA solution (\ref{Psi}):%
\begin{equation}
\Psi\left(  x,z\right)  =\left[  A+\varepsilon F_{\lambda}\left(  x\right)
e^{i\lambda z}+\varepsilon G_{\lambda}^{\ast}\left(  x\right)  e^{-i\lambda
^{\ast}z}\right]  e^{i\Xi\left(  x,z\right)  },\label{PP}%
\end{equation}
where the asterisk stands for the complex conjugation and $\varepsilon$ is a
real infinitesimal amplitude of the perturbation with complex eigenfunctions
$F_{\lambda}\left(  x\right)  $ and $G_{\lambda}\left(  x\right)  $, that are
associated to a complex eigenvalue, $\lambda$. As usual, the imaginary part of
$\lambda$ measures the instability growth rate of the perturbation, and thus
determines whether the CA solution is stable. The substitution of expression
(\ref{PP}) in Eq. (\ref{Model}) and linearization leads to the eigenvalue
problem in the matrix form
\begin{equation}
\left(
\begin{array}
[c]{cc}%
L_{1} & gA^{2}\\
-gA^{2} & L_{2}%
\end{array}
\right)  \left(
\begin{array}
[c]{c}%
F_{\lambda}\left(  x\right)  \\
G_{\lambda}\left(  x\right)
\end{array}
\right)  =\lambda\left(
\begin{array}
[c]{c}%
F_{\lambda}\left(  x\right)  \\
G_{\lambda}\left(  x\right)
\end{array}
\right)  ,\label{Eigenfunction}%
\end{equation}
where the operators $L_{1}$ and $L_{2}$ are
\begin{eqnarray}
L_{1} &=&gA^{2}+\frac{1}{2}\partial _{xx}+i\frac{\beta}{6} \partial _{xxx}+iW\partial _{x}
\notag \\
&&-\frac{\beta}{2} W_{x}\partial _{x}-\frac{\beta}{2} W\partial _{xx}-i\frac{\beta}{2} W^{2}\partial
_{x},  \label{L1} \\
L_{2} &=&-gA^{2}-\frac{1}{2}\partial _{xx}+i\frac{\beta}{6} \partial _{xxx}+iW\partial _{x}
\notag \\
&&+\frac{\beta}{2} W_{x}\partial _{x}+\frac{\beta}{2} W\partial _{xx}-i\frac{\beta}{2} W^{2}\partial
_{x}.  \label{L2}
\end{eqnarray}%

As said above, we apply the plane-wave-expansion method \cite{NatureCommunication6_7257} to
study the MI of the CA solution (\ref{Psi}) in the periodic $\mathcal{PT}%
$-symmetric potential $V(x)$ given by Eqs. (\ref{Vr}) and (\ref{Vi}). In the
framework of the method, when $W(x)$ is a periodic function with period
$2\pi/q$, perturbation eigenmodes $F_{\lambda}\left(  x\right)  $ and
$G_{\lambda}\left(  x\right)  $, along with $W(x)$ itself, can be expanded into
Fourier series, according to the Floquet-Bloch theorem:
\begin{align}
\left(
\begin{array}
[c]{c}%
F_{\lambda}\left(  x\right)  \\
G_{\lambda}\left(  x\right)
\end{array}
\right)   &  =%
{\displaystyle\sum_{n=-\infty}^{+\infty}}
\left(
\begin{array}
[c]{c}%
u_{n}(k)\\
v_{n}(k)
\end{array}
\right)  e^{i(qn+k)x},\label{FG_pw}\\
W(x) &  =%
{\displaystyle\sum_{n=-\infty}^{+\infty}}
W_{n}e^{iqnx},\label{W_pw}%
\end{align}
where $k$ is the Bloch momentum, the presence of which makes the eigenmodes
quasiperiodic. Substituting Eqs. (\ref{Vr}), (\ref{Vi}), (\ref{FG_pw}) and
(\ref{W_pw}) into the eigenvalue problem (\ref{Eigenfunction}), one arrives at
the following system of linear equations for perturbation amplitudes $u_{n}%
$, $v_{n}$ and eigenvalue $\lambda(k)$:
\begin{align}
\lambda u_{n} &  =gA^{2}v_{n}+\alpha_{n}u_{n}+\beta_{n-1}u_{n-1}\nonumber\\
&  +\gamma_{n+1}u_{n+1}+\theta_{n-2}u_{n-2}+\theta_{n+2}u_{n+2},\label{un}\\
\lambda v_{n} &  =-gA^{2}u_{n}+\sigma_{n}v_{n}+\delta_{n-1}v_{n-1}\nonumber\\
&  +\eta_{n+1}v_{n+1}+\theta_{n-2}v_{n-2}+\theta_{n+2}v_{n+2},\label{vn}%
\end{align}
where we define%
\begin{align*}
\alpha_{n} &  =gA^{2}-\frac{V_{0}}{8}\left(  4-\beta V_{0}\right)  \left(
qn+k\right)  \\
&  -\frac{1}{4}\left(  2-\beta V_{0}\right)  \left(  qn+k\right)  ^{2}%
+\frac{\beta}{6}\left(  qn+k\right)  ^{3},\\
\beta_{n} &  =-\frac{V_{1}}{4}\left(  2-\beta q\right)  \left(  qn+k\right)
+\frac{\beta V_{1}}{4}\left(  qn+k\right)  ^{2},\\
\gamma_{n} &  =-\frac{V_{1}}{4}\left(  2+\beta q\right)  \left(  qn+k\right)
+\frac{\beta V_{1}}{4}\left(  qn+k\right)  ^{2},\\
\theta_{n} &  =\frac{\beta V_{1}^{2}}{8}\left(  qn+k\right)  ,\\
\sigma_{n} &  =-gA^{2}-\frac{1}{8}V_{0}\left(  4-\beta V_{0}\right)  \left(
qn+k\right)  \\
&  +\frac{1}{4}\left(  2-\beta V_{0}\right)  \left(  qn+k\right)  ^{2}%
+\frac{\beta}{6}\left(  qn+k\right)  ^{3},\\
\delta_{n} &  =-\frac{V_{1}}{4}\left(  2+\beta q\right)  \left(  qn+k\right)
-\frac{\beta V_{1}}{4}\left(  qn+k\right)  ^{2},\\
\eta_{n} &  =-\frac{V_{1}}{4}\left(  2-\beta q\right)  \left(  qn+k\right)
-\frac{\beta V_{1}}{4}\left(  qn+k\right)  ^{2}.
\end{align*}

In the absence of the TOD ($\beta=0$) the results for the MI, i.e.,
eigenvalues $\lambda(k)$, produced by the present analysis, are consistent with those
reported in Ref. \cite{NatureCommunication6_7257}. In the following, the
growth rate of the MI of the CA solution is defined as the largest imaginary
part of $\lambda(k)$. As a typical example, Fig. 2 depicts the dependence of
$\max[\operatorname{Im}\{\lambda\}]$ on $k$ (in the half of the first
Brillouin zone) for the different TOD strengths (of either sign) in both the
self-focusing and defocusing regimes.

\begin{figure}[tbp]
\resizebox{0.48\textwidth}{!}{  \includegraphics{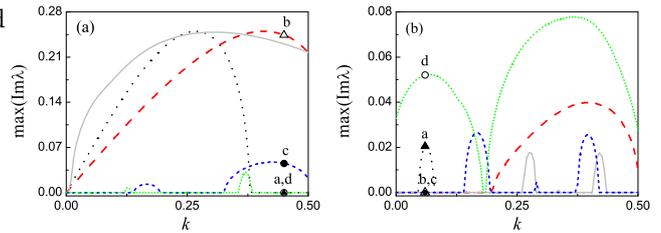}
} 
\caption{ The dependence of the MI growth rate, i.e., the largest imaginary
part of eigenvalues $\protect\lambda(k)$, on the Bloch wavenumber $k$ (in
the half of the first Brillouin zone) for different values of the TOD
coefficient, in the system containing periodic potential (\protect\ref{W}).
Panels (a) and (b) report the results, severally, for the self-focusing ($%
g=1 $) and defocusing ($g=-1$) nonlinearity, with the black dotted, red
dashed, gray solid, blue short-dashed, and green short-dotted curves
corresponding, respectively, to $\protect\beta=-3$, $-1$, $0$, $1$, and $3$
in (a), and $\protect\beta=-3$, $-1$, $0$, $0.2$, and $0.45$ in (b); note
the difference in the scales of vertical axes in panels (a) and (b). Here
the amplitude of the unperturbed solution is $A=0.5$, and the other
parameters are $V_{0}=4$, $V_{1}=0.2$, and $q=1$.}
\label{fig:2}
\end{figure}

For the self-focusing nonlinearity [Fig. 2(a)], at $\beta=0$, as well as at a
close value, $\beta=-1$, the MI eigenvalues with the largest imaginary part
are complex at all $k$ [see the gray solid and red dashed curves in Fig.
2(a)], hence the CA waves are linearly unstable to \emph{all perturbations},
as shown in Ref. \cite{NatureCommunication6_7257} for $\beta=0$. The situation
changes in the presence of sufficiently strong TOD. As shown by the black
dotted curve pertaining to $\beta=-3$ in Fig. 2(a), the CA solution is stable
against perturbations corresponding to the Floquet-Bloch modes close to the
edge of the Brillouin zone. The positive TOD, with $\beta>0$, strongly
suppresses the instability [see the blue and green short-dashed lines in Fig.
2(a), corresponding to $\beta=+1$ and $+3$, respectively]. Thus, the TOD terms
provides for partial stabilization of the CA waves, under the action of the
self-focusing nonlinearity. This finding may be qualitative understood as a manifestation of the trend,
imposed by the TOD, to replace the strong absolute instability by its weaker
convective counterpart.

The situation is different for the defocusing nonlinearity [Fig. 2(b)]. At
$\beta=0$, there are narrow MI bands \cite{NatureCommunication6_7257}, shown
by the gray solid curve in Fig. 2(b). Indeed, it is natural that the
self-defocusing gives rise to weaker MI, if any. The TOD with $\beta<0$
originally enhances the MI [see the red dashed curve in Fig. 2(b)
corresponding to $\beta=-1$], which is followed by the suppression of the MI
at larger values of $-\beta$, as shown by the black dotted curve corresponding
to $\beta=-3$. The MI is completely absent at $\beta<-3$, as can be seen below in
Fig. 6(b). On the other hand, the increase of $\beta>0$ leads to strong
amplification of the MI; in particular, the green short-dotted curve in Fig.
2(b), corresponding to a relatively small positive TOD coefficient,
$\beta=0.45$, demonstrates strong instability to the perturbation with
\emph{all values} of $k$.

\begin{figure}[tbp]
\resizebox{0.52\textwidth}{!}{  \includegraphics{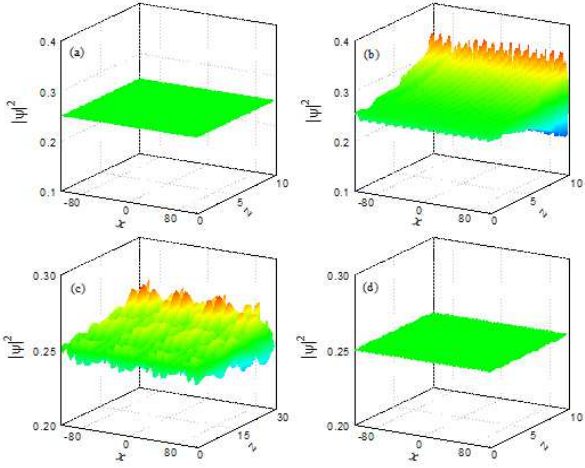}
} 
\caption{ Numerically simulated evolution of an unstable CA wave under the
action of the self-focusing nonlinearity ($g=1$) and periodic complex
potential (\protect\ref{W}). (a) $\protect\beta=-3;$ (b) $\protect\beta=-1$;
(c) $\protect\beta=1$, and (d) $\protect\beta=3$. Here, we fix the
perturbation parameters as $\protect\varepsilon=0.01$ and $k=0.45$, the
other parameters being the same as in Fig. 2. }
\label{fig:3}
\end{figure}

To explore the nonlinear development of the MI, we have performed systematic
simulations of Eq. (\ref{Model}), taking inputs in the form of unstable CA
solutions with the addition of small perturbations corresponding to specific
Floquet-Bloch eigenmodes, as per Eq. (\ref{PP}). The results are summarized in
Figs. 3 and 4, for the amplitude of the unperturbed solution $A=0.5$, and the
perturbation amplitude $\varepsilon=0.01$.

Figure 3 shows the evolution of the CA solution perturbed by the eigenmodes at
$k=0.45$ for different values of the TOD coefficient, $\beta$, in the
self-focusing regime. The so built solution is stable for $\beta=-3$ and $+3$
[Figs. 3(a,d)], and unstable for $\beta=-1$ and $+1$ [Figs. 3(b,c)]. These
findings are consistent with results of the linear-stability analysis, which
predicts no instability for $k=0.45$ and $\beta=\mp3$, as shown by the points
``a" and ``d" in Fig. 2(a) [although points ``a" and ``d" seem identical
in Fig. 2(a), they correspond to the different values of the TOD coefficient],
while for $\beta=-1$ and $+1$ the instability growth rates, i.e., $\max
[\operatorname{Im}\{\lambda\}]$, are $0.24428$ and $0.04544$, respectively,
see the points ``b" and ``c" in Fig. 2(a).

Figure 4(a) shows that, at negative $\beta$, the unstable CA wave is split by
the MI into a chain of soliton, which is the same outcome as in the case of
the usual self-focusing NLS equation with $\beta=0$, cf. Fig. 4(b). At
$\beta>0$, the MI cannot form a chain of solitons because the initial
perturbation eigenmode is not a simple periodic wave like sine or cosine, see
Fig. 4(c).

\begin{figure}[tbp]
\resizebox{0.48\textwidth}{!}{  \includegraphics{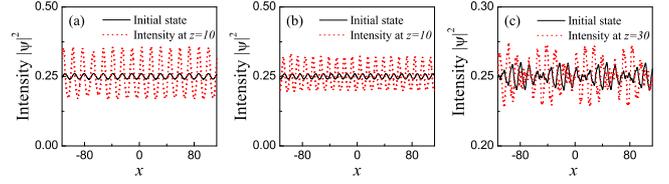}
} 
\caption{ The distributions of the initial intensity and the intensity at
the distance $z$, as produced by numerical simulations. (a) $\protect\beta%
=-1 $ and $z=10$; (b) $\protect\beta=0$ and $z=10$, and (c) $\protect\beta=1$
and $z=30$. Parameters are the same as in Fig. 3.}
\label{fig:4}
\end{figure}

Figure 5 displays the simulated evolution in the system with the defocusing
nonlinearity, where the perturbation eigenmodes with $k=0.06$ are initially
added. This figure demonstrates, in agreement with the prediction of the
linear-stability analysis summarized in Fig. 2(b), that the CA wave is stable
to these perturbations at $\beta=-1$ and $+0.2$, see Figs. 5(b) and 5(c) and
the corresponding points ``b" and
``c" in Fig. 2(b) [points ``b" and ``c", which seem identical in Fig. 2(b),
correspond to different values of the TOD coefficient].
On the other hand, the solution is unstable at
$\beta=0.45$, see Fig. 5(d) and the corresponding point ``d" in Fig. 2(b).
A specific situation is observed at
$\beta=-3$. While the linear-stability analysis predicts the MI in this case
[see the point ``a" in Fig. 2(b)], direct simulations reveal some fluctuations, but no exponentially growing
instability, as shown in Fig. 5(a). This happens because the corresponding
growth rate is very small, $\max[\operatorname{Im}\{\lambda\}]\approx$
$0.02049$. The result shows an example of \emph{weak stability} observed at
very small MI growth rates.

\begin{figure}[tbp]
\resizebox{0.52\textwidth}{!}{  \includegraphics{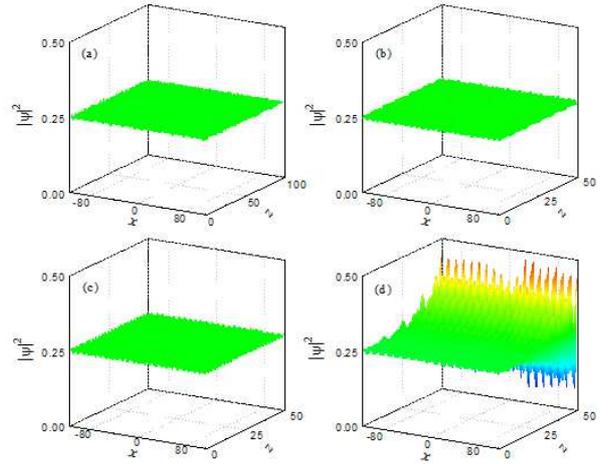}
} 
\caption{The same as in Fig. 3, but for the defocusing nonlinearity ($g=-1$%
), $\protect\varepsilon=0.01$, and $k=0.06$. (a) $\protect\beta=-3$; (b) $%
\protect\beta=-1$; (c) $\protect\beta=0.2$, and (d) $\protect\beta=0.45$.}
\label{fig:5}
\end{figure}

\begin{figure}[tbp]
\resizebox{0.49\textwidth}{!}{  \includegraphics{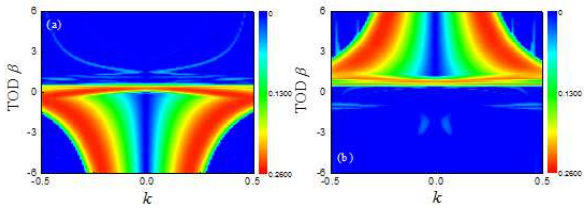}
} 
\caption{The dependence of the largest imaginary part of the MI eigenvalues
on the TOD strength, $\protect\beta$, in the first Brillouin zone of
periodic potential (\protect\ref{W}) for (a) $g=1$ and (b) $g=-1$
(self-focusing and defocusing, respectively), while the other parameters are
the same as in Fig. 2. }
\label{fig:6}
\end{figure}

To display the effect of the TOD on the MI of the CA waves, Fig. 6 depicts the
dependence of the largest imaginary part of the MI eigenvalues on the TOD
coefficient, $\beta$, in the first Brillouin zone for the self-focusing and
defocusing nonlinearities. For the self-focusing case, Fig. 6(a) shows that
the MI takes place (with complex most unstable eigenvalues) for all $k$ when
$\beta$ belongs to interval $-1.6<\beta<0.66$. However, at $\beta<-1.6$, a
stability band emerges at the edge of the Brillouin zone, whose size grows
with the increase of $|\beta|$, while as $\beta>0.66$, the stability band is
quite broad, while the remaining instability is weak.

Figure 6(b) shows an opposite situation in the case of the defocusing
nonlinearity. The difference is explained by the fact that, at $\beta=0$, a
stability band does not exist for the self-focusing nonlinearity, but it is
present in the defocusing case, as can be seen in Fig. 2(b). Despite the
difference, a general conclusion is that the TOD effect eventually tends to
attenuate the MI of the CA waves.

\begin{figure}[tbp]
\resizebox{0.49\textwidth}{!}{  \includegraphics{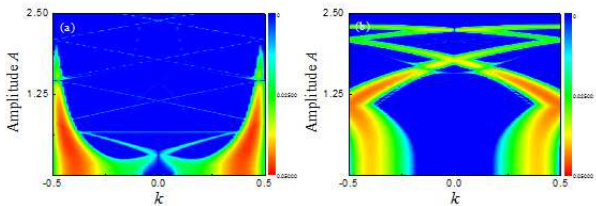}
} 
\caption{The dependence of the MI growth rate, on the amplitude, $A$, and
wavenumber, $k$, of the CA solution, supported by the complex periodic
potential (\protect\ref{W}). (a) $g=1$, $\protect\beta=1$; (b) $g=-1$, $%
\protect\beta=-1$, the other parameters being $V_{0}=4$, $V_{1}=0.2$, and $%
q=1$.  }
\label{fig:7}
\end{figure}

The above analysis was reported for the MI of the CA solution with $A=0.5$.
Now, we aim to consider the effect of the variation of the amplitude on the
MI. Figure 7 presents the dependence of the MI growth rate on $A$ in the first
Brillouin zone for the self-focusing and defocusing nonlinearities,
respectively. Fig. 7(a) demonstrates that, for the self-focusing nonlinearity,
the instability bands are mainly concentrated at the edge of the first
Brillouin zone for $\beta=1$, shrinking with the increase of $A$ until
reaching $A\approx2$, where the MI virtually disappears. For the defocusing
nonlinearity, Fig. 7(b) demonstrates that the stability band is mainly focused
at the center of the first Brillouin zone. Its size grows with the increase of
$A$ up to $A\approx1.1$, at $\beta=-1$. With the further increasing of $A$,
the instability bands start to converge and cross until vanishing at
$A\approx2.3$.

\begin{figure}[tbp]
\resizebox{0.49\textwidth}{!}{  \includegraphics{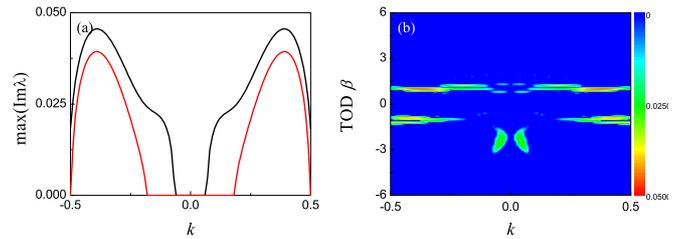}
} 
\caption{(a) The dependence of the largest imaginary part of eigenvalues on
the Bloch wavenumber $k$ for the zero state, which corresponds to the zero
amplitude of the underlying CA solution, $A=0$. Here, the black solid and
red solid curves represent $g=1$, $\protect\beta =1$ and $g=-1$, $\protect%
\beta =-1$, respectively. (b) The dependence of the MI growth rate on the
TOD strength $\protect\beta $ and wavenumber $k$ for the self-focusing
nonlinearity ($g=1$). The other parameters are the same as in Fig. \protect\ref%
{fig:7}.}
\label{fig:8}
\end{figure}

Lastly, it is relevant to address the stability of the zero state in the
model based on Eq. (\ref{Model}), which corresponds to amplitude $A=0$ of
the unperturbed CA state. Usually, the stability of the zero background is
an important issue in the analysis of $\mathcal{PT}$-symmetric systems \cite%
{Rep.Prog.Phys70_947,PRL80_5243,JMP43_2814,PRL89_270401}. From Fig. \ref{fig:8}, one can
conclude that the stability bands of the zero state are mainly concentrated
at the center of the first Brillouin zone for both the self-focusing and
defocusing nonlinearity, as shown in Fig. \ref{fig:8}(a), in which, at $g=1$,
the stability band of the zero solution is $|k|\leq 0.06$ for $\beta =1$,
while at $g=-1$ it is $|k|\leq 0.18$ for $\beta =-1$. In this case, the
dependence of the instability growth rate of the zero solution on the TOD
strength $\beta $ and wavenumber $k$ is shown in Fig. \ref{fig:8}(b). From
here, one can see that the zero state is stable for all $k$ when $\beta $
belongs to intervals $\beta \leq -3.3$, or $-0.8<\beta <0.6$, or $\beta >2$.
At $-1.5<\beta <-3.3$, the instability band is mainly concentrated near the
center of the first Brillouin zone, while, at $-1.5<\beta <-0.8$, the
instability is concentrated at the edge of the first Brillouin zone. For $%
0.6<\beta <2$, instability bandgaps form at some values of $k$.

\section{Conclusion}

We have considered the CA (constant-amplitude) waves governed by the NLS
equation\ which includes the spatial TOD (third-order dispersion) and
complex potentials (thus corresponding to a non-Hermitian Hamiltonian). The
model was built as a solution of the inverse problem, which predicts the
complex potential needed to support a CA state with a required
phase-gradient structure. The setting can be realized in optical nonlinear
waveguides, with an appropriate distribution of the local gain and loss. The
results show, first, that the diffraction of the truncated CA solution is
strongly suppressed, if the state carries the correct phase structure. The
main part of the analysis was focused on the MI (modulation instability) of
the CA waves, by means of the linear-stability analysis based on the
plane-wave expansion method, and with the help of direct numerical
simulations. The results have shown that the TOD tends to attenuate the MI.
This may be understood as a manifestation of the trend, imposed by the TOD,
to transform strong absolute instability into a weaker convective form.
Direct simulations of the perturbed evolution of the CA waves have revealed
a phenomenon of the weak stability, which occurs at sufficiently small
values of the MI instability growth rate, when formally unstable CAs turn
out to be effectively stable, in terms of the simulated evolution. The
approach elaborated in the present work can be used to analyze the stability
band structure of periodic solutions in periodic complex potentials for
other wave systems.

\section*{Acknowledgments}

This research is supported by the National Natural Science Foundation of China
grant 61078079 and 61475198, the Shanxi Scholarship Council of China grant
2011-010 and 2015-011. The work of B.A.M. is supported, in part, by the joint
program in physics between the National Science Foundation (US) and Binational
Science Foundation (US-Israel), through Grant No. 2015616.

\end{document}